\documentclass[journal]{IEEEtran}

\usepackage{myStyleIEEE}
\usepackage{nomencl}
\usepackage{ragged2e}
\usepackage{graphicx}
\usepackage{makecell}

\IEEEoverridecommandlockouts

\newlength{\bracewidth}

\usepackage{etoolbox}
\renewcommand\nomgroup[1]{%
  \item[\bfseries
  \ifstrequal{#1}{P}{A. Parameters}{%
  \ifstrequal{#1}{V}{C. Variables}{%
  \ifstrequal{#1}{S}{B. Sets and Indices}{}}}%
]}

\begin{document}
% \bstctlcite{IEEEexample:BSTcontrol}

\title{Control-mode as a Grid Service in Software-defined Power Grids: GFL vs GFM}

\newtheorem{proposition}{Proposition}
\renewcommand{\theenumi}{\alph{enumi}}

\author{Guoxuan~Cui,~\IEEEmembership{Student~Member,~IEEE,} 
        Zhongda~Chu,~\IEEEmembership{Member,~IEEE,} and
        Fei~Teng,~\IEEEmembership{Senior Member,~IEEE} 
        % \thanks{Zhongda Chu and Fei Teng (e-mail: f.teng@imperial.ac.uk) are with Imperial College London.
        
        % <-this % stops a space
        
\vspace{-0.5cm}}
\maketitle
\IEEEpeerreviewmaketitle

%ABSTRACT
\begin{abstract}
In power systems with high penetration of power electronics, grid-forming control is proposed to replace traditional Grid-Following Converter (GFL) in order to improve the overall system strength and resist small-signal instability in weak grids by directly forming the terminal voltage. However, sufficient headroom of both active and reactive power must be made available for Grid-Forming Converter (GFM) to operate, potentially leading to sub-optimal operation in steady states. This presents a new research problem to optimally allocate between GFM and GFL to balance the ability of GFMs to improve the grid strength and the potential economic loss resulting from reserved headroom. An optimization framework under software-defined grids is proposed, for the first time, to dynamically determine the optimal allocation of GFMs and GFLs in power systems at each time step of system scheduling according to system conditions, which ensures both system stability and minimum operational cost. To achieve this, the system scheduling model is expanded to simultaneously consider the constraints related to active and reactive power reserves for GFMs, as well as the system level stability. Case studies conducted on the modified IEEE 30-bus system demonstrate significant economic benefits in that the optimal proportion of GFMs in the power system can be dynamically determined while ensuring power reserve and grid stability constraints. 
\end{abstract}


\makenomenclature
\renewcommand{\nomname}{List of Symbols}
\mbox{}
\nomenclature[P]{$M$}{system inertia$\,[\mathrm{MWs/Hz}]$}
% \printnomenclature

\section{Introduction} \label{sec:1}
Decarbonization targets have been set by many countries recently, to reduce their dependence on fossil fuels such as coal and gas. In Europe, the process was significantly accelerated in 2022 due to rising gas prices. This transition from traditional non-renewable fuels to modern Renewable Energy Sources (RES) involves the gradual replacement of conventional Synchronous Generators (SGs) with power electronic converters. However, power systems with high RES penetration encounter new challenges that can compromise their stable operation. These challenges include frequency and voltage control, system inertia management, and fault ride-through capability.

Grid-connected power converters, which do not contribute the rotational inertia to the external grid, are widely used in Type III and Type IV wind turbines and PV solar power generation \cite{7866938}. Most grid-connected converters currently operate under grid-following control mode, which programs the converter to act as a current source to provide the maximum power \cite{8070502}. However, high penetration of GFLs that use Phase-locked Loops (PLL) for synchronization with the external grid can lead to instability or loss of synchronization in weak power systems \cite{zhang2021grid,8743441,9714816}. Extensive research has demonstrated that GFLs utilizing PLL for synchronization can cause small-signal instability issues \cite{7079502,7027822}. To address this issue, the grid-forming control algorithm has been proposed to actively regulate converters' terminal frequency and voltage by emulating the dynamics of traditional SGs \cite{8586162}. GFMs have been shown to be effective in weak grids \cite{8973583} and can also enhance system strength by operating as a voltage source behind an impedance \cite{9714816,8488538}. As a result, the higher the GFMs penetration in power systems, the stronger the grid becomes.

On the other hand, 
% conventional generators can supply frequency regulation services through their inertia and damping contributions during the loss of generation or load events, as well as voltage support services by injecting reactive current during short circuit events. 
with the graduate retirement of conventional generators, system operators are increasingly requiring renewable power plants to provide ancillary services. 
% However, the converters are initially designed to synchronize the frequency between the rotor and the external grid and maximize output power. Therefore, enabling converter-interfaced RES to provide ancillary services similar to conventional generators requires further attention due to the physical limits of power electronic components. 
%The grid-connected inverter was initially designed for variable-speed wind turbines to convert variable frequency, voltage, and power generated by the rotor into constant frequency and power while extracting maximum power output from available wind power \cite{1331485}. 
% Doubly-Fed Induction Generators (DFIG) are commonly used in wind turbines and synchronization is typically achieved through a PLL, which the fundamental control loop was illustrated in \cite{844502}. 
Therefore, extensive research has been conducted on designing additional control loops for GFLs and GFMs to achieve more functionality.
Frequency and voltage regulators can be employed to operate GFL at grid-supporting mode \cite{ZUO2021100496,6200347,9420359}, whilst the grid-following virtual inertia device is developed in \cite{8607122} to mimic the dynamic of SGs and provide fast frequency response.
% Active power is regulated by $P-f$ droop control when a frequency deviation is detected. Likewise, the $Q-v$ droop control loop supports the node voltage during a fault by increasing the reactive current flow through the converter into the grid. 
% Moreover, grid-following virtual inertia device is developed in \cite{8607122} to mimic the dynamic of SGs and provide fast frequency response. 
However, one benefit of GFL is its straightforward control structure \cite{9589432}, and introducing redundant control loops may increase the complexity of the GFL, potentially leading to unforeseen errors.
% Furthermore, existing literature and results indicate that the grid-forming control strategy outperforms the grid-following one by achieving better frequency containment and a lower relative Rate of Change of Frequency (RoCoF) \cite{ZUO2021100496}. 
Furthermore, the simulation results from \cite{8607122} illustrate that while both GFL and GFM virtual inertia implementations improve the frequency nadir and maximum Rate of Change of Frequency (RoCoF), GFM performs better in terms of absolute values and requires less total inertia and damping. In addition, GFM achieves a smaller maximum power injection in response to a frequency event with less control effort compared to GFL.
% During a short circuit fault, the GB Grid Code requires power electronic devices to provide voltage support services \cite{gridcode}. However, fast fault current injection is listed separately as one of the capacities of GFMs, which is more strict and effective compared to the requirements for GFLs. The GB Grid Code \cite{gridcode} indicates that the reactive current injection of GFMs should respond 4 times faster than that of GFLs. 
As for the voltage support, GFM can directly control terminal voltage, resulting in a fast and accurate response to a short circuit fault, while GFL may require an additional $Q-v$ droop control loop for the voltage regulation. It can be suggested that GFM may achieve better results than GFL in both frequency regulation and voltage support services. Therefore, with the support of GFMs in the network, GFL can be considered a grid-feeding converter that only tracks the MPPT point and injects active power into the grid based on the power reference without providing any ancillary services.

% Studies \cite{xin2023gridforming,9282199} have considered the small-signal stability margin of a complex system with both GFL and GFM operating actively but do not specify the upper boundary for the penetration of GFMs in a power system. 
While it is clear that GFMs will play a crucial role in future power systems, the potential drawbacks of excessive GFM penetration must also be taken into account. In contrast to GFLs, additional active and reactive power headroom is required for GFMs to provide grid services. For instance, grid-forming controlled wind turbines need to operate at a sub-optimal point, meaning sufficient energy reserve or power headroom in steady states must be considered at a conservative level to guarantee support for the external system under any contingencies. The deloading control technique is commonly utilized to maintain a power reserve margin by adjusting the optimal power extraction point to a lower level \cite{DREIDY2017144}, leading to potential economic loss. Additionally, in terms of system stability issues, \cite{9714816} illustrates that the GFMs can become unstable when the grid is strong with a small grid impedance $Z_g$. 
% Nevertheless, the majority of research only considers the deloaded power as a fixed value (e.g., 10\% of its rated capacity) and fails to achieve global optimality. As a result, the optimal allocation of GFM to achieve cost-effective system scheduling while ensuring system stability remains an open question.

As a result, the placement and allocation of GFMs for cost-effective system operation while ensuring system stability become an emerging research topic in recent years. One study \cite{8607122} optimizes the parameters and location of GFL and GFM virtual inertia devices in a power system, using system norms to assess transient stability with inertia and damping constraints. However, this study only considers transient frequency stability analysis and does not address the small signal instability issue imposed by PLL. Other studies \cite{xin2023gridforming, 9282199} focus on the optimal GFM penetration and allocation problem from the grid strength perspective, without considering the frequency or voltage regulation services and corresponding operational constraints of GFM. Furthermore, existing research only considers the optimal allocation of GFMs during the system planning level, assuming the control mode will keep constant in the operation time scale. However, this can result in sub-optimal performance, as the real-time system operation point and the need for services vary significantly.

Due to the advancements in information and communication technology, Software-Defined Networking (SDN) has become increasingly popular in recent years. The prominent feature of SDN is the separation of the control and data planes, which enhances the flexibility and programmability of the network \cite{6834762}. Building on this concept, the software-defined grid has been proposed as a means of achieving fast and centralized global optimality, leveraging the advancements in communication systems of future power networks \cite{Software-defined-grid}. Benefiting from the software-defined grid, dynamic adjustment and optimization of the control algorithm, grid topology, and scheduling strategy of a system through software is achievable. According to \cite{8525299}, the control strategy and parameters of Inverter-Based Resources (IBR) can be determined through predefined logic or system-level commands with fast communication, which enables the dynamical allocation of GFL and GFM at the system scheduling level.

The primary objective of this paper is to develop a dynamical optimization model to determine  GFM allocation in software-defined girds such that the efficiency loss due to the reserving energy and headroom for frequency regulation, phase jump response and voltage support under normal operation can be minimized while maintaining power system frequency and small-signal stability.
% derive the dynamically changing optimal penetration of GFMs while considering the power generation loss caused by reserving headroom for frequency regulation, phase jump response and voltage support under normal operating conditions. This is achieved while ensuring the small-signal stability of the power system. 
The key contributions include:

\begin{itemize}  %[\hspace{1em}1)]
    \item An optimization framework for software-defined grids is proposed to dynamically determine the optimal allocation of GFMs in power systems. The framework is integrated into the Stochastic Unit Commitment (SUC) problem to optimize operating conditions and inform cost-effective scheduling on an hourly/half-hourly basis.
    
    % The framework takes into account the steady-state power generation deloading loss of GFMs, which is necessary for providing active or reactive power support during unexpected events. Additionally, the algorithm leverages the exceptional ability of GFMs in improving grid strength and mitigating small-signal instability. 

    \item The constraints for GFM power reserves are derived analytically by combining the active power reserve for frequency regulation and phase jump power injection services with the reactive power reserve for voltage support during transient fault events. Moreover, the Small-signal Stability Constraint (SSC) is formulated based on the operation status of SGs and the penetration level of GFMs. These GFM capability-related constraints are incorporated together into the system scheduling model for the first time.
    
    % \item The Small-signal Stability Constraint (SSC) is formulated based on the operational status of SGs and the penetration level of GFMs in each individual wind farm. By combining the SSC with the power reserve constraints of GFMs, an optimal solution can be obtained for the trade-off problem of determining the exact amount of GFM required from a system scheduling perspective.
    
    \item The advantages and benefits of the proposed optimization framework in software-defined grids for determining the penetration of GFM are demonstrated through case studies based on a modified IEEE 30-bus system. The validation of the derived power reserve constraints and SSC is provided, and the cost-effectiveness of the advanced approach is fully illustrated.
\end{itemize}

The remainder of the paper is structured as follows. Section II presents GFM operational constraints, including frequency regulation, phase jump response, and reactive power support services by GFMs. Section III derives the GFM-related power system stability constraints, taking into account the frequency nadir and RoCoF constraints, as well as the formulation and linearization of the SSC. Section IV outlines the overall structure of the SUC problem, which minimizes the total cost of the power system while maintaining system stability. The performance and improvements of the proposed algorithm are demonstrated through case studies in Section V. Finally, Section VI concludes the paper.

\section{Operational Constraints for GFMs} \label{sec:2}
% In the event of unplanned occurrences, frequency stability constraints such as frequency nadir and RoCoF guarantee a stable system operation by injecting or absorbing additional active power. The total inertia and power reserves from SGs and GFMs in the system must be such that these constraints are maintained during disturbances. Moreover, active power response is also triggered during a phase jump event. Therefore, the implementation of GFMs based on the deloading control technique results in active power reserve constraints of GFMs in steady states. Likewise, system faults causing voltage sag events necessitate reactive power injection to maintain terminal voltage, resulting in the reactive power reserve constraint for GFMs in steady states.

Based on the grid-forming capability guidelines published by National Grid Electricity System Operator Limited (NGESO) \cite{GFM_NGESO} and the European Network of Transmission System Operators for Electricity (ENTSO-E) \cite{GFM_ENTSOE}, the comparison of the general capability requirements between GFM and GFL is demonstrated in Table~\ref{tab:GFLvsGFM}. It can be observed that GFM is mandated to provide a greater number of services, including RoCoF response, active phase jump power provision, and fast fault current injection, in contrast to GFL. In case of frequency events, GFMs within the system are required to provide active power support for inertia contribution and frequency regulation services, ensuring that system security constraints are upheld during disruptions. Additionally, active power response is also triggered during phase jump events. As a result, the implementation of GFMs based on the deloading control technique imposes active power reserve constraints on GFMs in steady states. Similarly, system faults causing voltage sag events necessitate reactive power injection to maintain terminal voltage, leading to the reactive power reserve constraint for GFMs in steady states. 
The derivation of the operational constraints to facilitate GFM capabilities, as previously discussed, is presented in this section.

\begin{table}[t]
\renewcommand{\arraystretch}{1.4}
\caption{A comparison of GFM/GFL capability requirements}
\label{tab:GFLvsGFM}
\noindent
\centering
\resizebox{\columnwidth}{!}{%
\begin{minipage}{\linewidth}
 \renewcommand\footnoterule{\vspace*{-5pt}}
\begin{center}
\begin{tabularx}{0.99\textwidth} {
         >{\raggedright\arraybackslash}X
        | >{\raggedright\arraybackslash}X
        | >{\raggedright\arraybackslash}X
        }
\Xhline{1px}
 &
    \textbf{GFL} &
    \textbf{GFM}
    \\ \hline
\textbf{Contributing to total system inertia }&
  No requirement in terms of inertia contribution. &
  Being required to supply active RoCoF response duration a frequency change event. \\ \hline
\textbf{Providing active phase jump power} &
  No requirement in terms of phase jump power provision. &
  Being expected to inject/absorb active phase jump power as a result of phase angle changes. \\ \hline
\textbf{Contributing to fault level (Short circuit power)} &
  No specific requirement in terms of fault current injection. &
  Being required to inject reactive current under fault conditions. \\
    \Xhline{1px}
\end{tabularx}

\end{center}
\end{minipage}
}
        \vspace{-0.4cm}
\end{table}

\subsection{Active Power Support Constraints}
\subsubsection{System Frequency Support}
During steady states, GFMs provide active and reactive power to the system in accordance with AC optimal power flow. To meet the stringent constraints of maximum RoCoF and frequency nadir, the necessary active power injection for each GFM wind turbine is determined based on the frequency dynamics in \cite{7115982} as: 
\begin{align}
\label{Delta_Pf}
            &\Delta P_{{f}_{n,j}} = 2{{H}_{\mathrm{GFM}_{n,j}}}\frac{\partial\Delta f(t)}{\partial t}, &  &\forall n \in \mathcal N, \forall j \in \mathcal F, 
\end{align}
where $\mathcal{N}$ represents the set of GFM wind turbines, with the number of GFM wind turbines denoted by $\left| \mathcal N \right| = N_\mathrm{GFM}$. ${H}_{\mathrm{GFM}_{n,j}}$ stands for the synthetic inertia provided by the $n$-th GFM wind turbine in the $j$-th wind farm with $\mathcal{F}$ being the set of wind farms, and $\Delta f(t)$ is the system frequency deviation. It should be noted that $\Delta P_{{f}_{n,j}}$ is positive in the event of an under-frequency. To ensure the system frequency security, the available wind power of GFM must be sufficient to provide frequency response service, resulting in:
\begin{equation}
\label{Pa}
            {P_{a_{n,j}}} -\Delta P_{{f}_{n,j}} - \Delta P^{\mathrm{GFM}}_\mathrm{WS_{n,j}} = P_{{0}_{n,j}} \ge 0.
\end{equation}
% In three-phase systems, the Park transformation allows for the representation of sinusoidal currents as DC values in a rotating dq frame that is synchronized with the detected grid fundamental frequency \cite{6739417}. 
% Consequently, $P_0$ and $Q_0$ denote the steady-state active and reactive power injection from GFM wind turbines into the grid, 
% which can be calculated as follows:
% \begin{equation}
% \label{ActivePower}
% P_{0} = {V_dI_{d0}+V_qI_{q0}},
% \end{equation}
% \begin{equation}
% \label{ReactivePower}
% Q_{0} = {V_dI_{q0}-V_qI_{d0}}.
% \end{equation}
% In steady-state, the d-axis and q-axis components of terminal voltage and current flowing through the converter are represented by $V_d$ and $I_{d0}$, and $V_q$ and $I_{q0}$, respectively. In addition, the d-axis of an individual IBR reference frame is aligned with the IBR terminal voltage in steady-state, resulting in the component of voltage on the q-axis $V_q=0$. The equations can be expressed as follows:
% \begin{equation}
% \label{ActivePower1}
% P_{0} = {V_eI_{d0}},
% \end{equation}
% \begin{equation}
% \label{ReactivePower1}
% Q_{0} = {V_eI_{q0}}.
% \end{equation}
In this constraint, $P_{a_{n,j}}$ represents the available wind power for a GFM wind turbine, while $\Delta P^{\mathrm{GFM}}_\mathrm{WS_{n,j}}$ denotes the wind shed by the GFM wind turbine. Understandably, $P_{{0}_{n,j}}$ denotes the steady-state active power injection from the GFM wind turbine into the grid. In this scenario, the wind turbine can operate under grid-forming control only if it satisfies the active power reserve constraint defined by equations \eqref{Delta_Pf} and \eqref{Pa}. The penetration of GFM is therefore closely related to the wind condition $P_{a_{n,j}}$ and the system inertia requirement $\Delta P_{{f}_{n,j}}$.

\subsubsection{Phase Jump Constraint}
\begin{figure}[b]
        \vspace{-0.4cm}
    \centering
	\scalebox{0.7}{\includegraphics[]{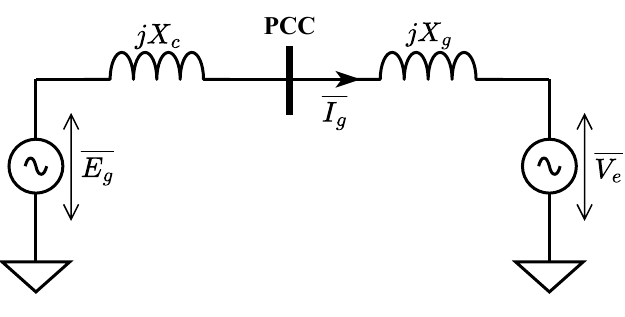}}
    \caption{\label{fig:EquivalentCircuit}Electrical equivalent circuit between the GFM and the external grid.} 

\end{figure}

The phase jump for a GFM refers to a shift in phase angle between the grid voltage of the system and the converter’s internal voltage. According to the commonly depicted GFM configuration, the electrical equivalent circuit between the converter and the grid can be observed in Fig.~\ref{fig:EquivalentCircuit}. 
In emulation of SGs, the active power that can be delivered by a GFM is expressed as follows:
\begin{equation}
\label{ActivePowerphase}
    P_{{0}_{n,j}} = \frac{E_gV_e \sin\delta_{n,j}}{X},
\end{equation}
where $E_g = \left| \overline{E_g} \right|$ represents the magnitude of the constant internal voltage source of GFM and $V_e = \left| \overline{V_e} \right|$ denotes the voltage magnitude of the external grid. The total reactance between GFM and the grid is represented by $X$, which is the sum of converter reactance $X_c$ and grid side reactance $X_g$. $\delta_{n,j}$ is defined as the phase difference between the $\overline{E_g}$ and $\overline{V_e}$. During a phase jump event, the angle difference will either increase or decrease depending on the real-world situation. Moreover, the active power will be delivered or absorbed instantaneously to prevent an asynchronization event from occurring.

According to the GB Grid Code, the grid-forming plant must have the ability to inject or absorb active power during a phase jump event\cite{gridcode}. The active phase jump power must be delivered immediately upon detection of a phase angle difference between the GFM internal voltage and the external grid voltage. The phase jump power is a natural response and its response time scale must be faster than that of either the active inertia power or the active damping power.

The phase jump angle is characterised as ${\Delta{\delta_{n,j}}}$ and according to the current GB Grid Code, the phase jump angle limit is ${\Delta{\delta}^\mathrm{max}} = {5^{\degree}}$ which ensures that the grid forming plant remains in normal operating condition without activating current limiting functions \cite{gridcode}. The paper primarily considers the power reserve requirements for GFM in steady states and therefore only takes into account the phase jump event that causes active power injection to the grid. The active phase jump power for the GFM wind turbine, represented by ${\Delta P_{{ph}_{n,j}}}$, is calculated as follows: 
\begin{equation}
\label{PhasejumpPower}
    \Delta P_{{ph}_{n,j}} = \frac{E_gV_e\left[\sin\left(\delta_{n,j}+\Delta\delta_{n,j}\right)-\sin\delta_{n,j}\right]}{X}.
\end{equation}
Therefore, to meet the requirement of GFM active phase jump response, GFMs must deload and reserve a conservative active power output of $\Delta P^\mathrm{max}_{{ph}}$ when ${\Delta{\delta}^\mathrm{max}}$ is applied in steady states instead of following the MPPT control scheme. Due to the relatively small magnitude of ${\Delta{\delta}^\mathrm{max}}$, the maximum active phase jump power can be derived as:
\begin{equation}
\label{PhasejumpPowerMax}
    \Delta P^\mathrm{max}_{{ph}} = \frac{E_gV_e\sin\left({\Delta{\delta}^\mathrm{max}}\right)}{X}.
\end{equation}
Considering the previously derived frequency stability-related active power constraint \eqref{Pa}, the updated constraint for a single GFM wind turbine is expressed as follows:
\begin{equation}
\label{Pa2}
{P_{a_{n,j}}} -\Delta P_{{f}_{n,j}} - \Delta P^\mathrm{max}_{{ph}} -  \Delta P^{\mathrm{GFM}}_\mathrm{WS_{n,j}} = P_{{0}_{n,j}} \ge 0.
\end{equation}
In addition, the capacity constraint of GFMs including the active and reactive power, should also be considered as follows:
\begin{equation}
\label{ActivePowerConstraints2}
    \sqrt{\left(P_{{0}_{n,j}}+\Delta P_{{f}_{n,j}}+\Delta P^\mathrm{max}_{{ph}}\right)^2+{Q_{{0}_{n,j}}}^2} \leq S_{\mathrm{GFM}_{n,j}},
\end{equation}
where ${Q_{{0}_{n,j}}}$ denotes the steady-state reactive power injection from GFM wind turbines into the grid, and $S_{\mathrm{GFM}_{n,j}}$ represents the rated capacity of $n$-th GFM wind turbine in the $j$-th wind farm. Unlike the frequency response, the phase jump power is associated with individual GFM wind turbines to address its own synchronization issues. Consequently, a higher GFM penetration level may necessitate a higher deloaded power to comply with the GFM ancillary services required by GB GFM Grid Code. 

As a result, \eqref{ActivePowerConstraints2} can be expressed as a SOC form and incorporated into the Mixed Integer Second-Order Cone Programming (MISOCP) based SUC problem. Under the condition of \eqref{Pa2} and \eqref{ActivePowerConstraints2}, the steady state active power output $P_{{0}_{n,j}}$ must be regulated to ensure that sufficient headroom is reserved for frequency and phase angle regulation, while reactive power output $Q_{{0}_{n,j}}$ remains at its optimal level to ensure compliance with voltage security constraints. Reactive power related transient constraints will be discussed in the next section along with current limitations.

\subsection{Reactive Power Support Constraints}
The GB Grid Code stipulates that nominal voltage variation at a specific node must remain within the range of $[0.95, 1.05]\,\mathrm{p.u.}$ during normal operation. Moreover, it is required that grid-forming plants must have the ability to provide reactive power in steady states based on grid conditions to maintain terminal voltage. In the event of a short circuit fault, a significant voltage drop may occur, causing severe grid instability. To mitigate this issue, GFM wind turbines must have the capability for fast fault reactive current injection, delivering reactive power into the system within 5ms of detecting a voltage sag event \cite{gridcode}. 
% However, the Grid Code indicates the reactive current injection is a function of post-fault transit voltage. 
% Due to the physical constraints of power electronic devices, the maximum reactive current is limited.

% \begin{figure}[!h]
%     \centering
% 	\scalebox{0.3}{\includegraphics[]{GFM_Figures/Reactive current.PNG}}
%     \caption{\label{fig:ReactiveCurrent}Reactive current injection level with respect to the faulted voltage condition.}\textcolor{blue}{redraw}
% \end{figure}

As a result, the reactive power reserve must be evaluated considering the GFM’s provision of reactive power in steady states and the large amount of reactive power injection required for voltage support service during transients. The system operator must ensure that the total power and current delivered by the GFM do not exceed its rated capacity. During a short circuit event, the additional reactive power injection $\Delta Q_{v_{n,j}}$ can be calculated based on the transient GFM terminal voltage $V_{t_{n,j}}$ and the required reactive current $\Delta I_{r_{n,j}}$ that flows from the converter to the grid:
\begin{equation}
\label{ReactivePowerInjection}
             \Delta Q_{v_{n,j}} = {V_{t_{n,j}}\Delta I_{r_{n,j}}},
\end{equation}
and $\Delta I_{r_{n,j}}$ must satisfy the requirements of the GB Grid Code, which specifies the relationship between transient voltage level and reactive current. 
% Fig.~\ref{fig:ReactiveCurrent} illustrates this correlation with different converter peak current ratings. 
The current limit is selected as $I_{\mathrm{limit}} = 1.5\, \mathrm{p.u.}$ for a superior voltage support ability and is also commonly used in practice. The reactive power injection can be expressed as: 
\begin{equation}
\label{ReactiveCurrentInjection}
             \Delta I_{r_{n,j}} = {d(V_{e_{n,j}}-V_{t_{n,j}})}.
\end{equation}
in which the droop gain of the reactive current control is $d = 0.6$, as per the given requirement. By substituting equation \eqref{ReactiveCurrentInjection} into equation \eqref{ReactivePowerInjection}, the relationship between reactive power injection and transient voltage can be derived:
\begin{equation}
\label{ReactivePowerInjectionGridCode}
             \Delta Q_{v_{n,j}} = {-dV_{t_{n,j}}^2+dV_{e_{n,j}}V_{t_{n,j}}}.
\end{equation}

Moreover, during a voltage sag event, the decrease in transient grid voltage leads to a reduction in the original active and reactive power output. As a result, the transient power outputs will be lower than the initial steady-state output and can be calculated using the fault voltage $V_{t_{n,j}}$: 
\begin{equation}
\label{ActivePowerTransient}
             P_{0_{n,j}}' = {V_{t_{n,j}}I_{a0_{n,j}}},
\end{equation}
\begin{equation}
\label{ReactivePowerTransient}
             Q_{0_{n,j}}' = {V_{t_{n,j}}I_{r0_{n,j}}},
\end{equation}
where $I_{a0_{n,j}}$ and $I_{r0_{n,j}}$ denote the active and reactive current in steady-state, respectively. During short circuit events, the requirement of reactive current $\Delta I_{r_{n,j}}$ and reactive power $\Delta Q_{v_{n,j}}$ causes a decrease in the priority of active power output. Consequently, the active current and power output can be decreased if there is not enough headroom for reactive current/power injection: 
\begin{equation}
\label{ActivePowerDrop}
             \Delta P_{v_{n,j}} = {V_{t_{n,j}}\Delta I_{a_{n,j}}}, (\Delta I_{a_{n,j}} \le 0, \Delta P_{v_{n,j}} \le 0),
\end{equation}
where $\Delta P_{v_{n,j}}$ represents the decrease in active power and is a function of $V_{t_{n,j}}$. Its magnitude should be determined based on the severity of the voltage sag event. The GFM power constraint for transient voltage support events can be derived by ensuring that the transient total power output does not exceed the rated capacity:
\begin{equation}
\label{ReactivePowerConstraints}
            \sqrt{(P_{0_{n,j}}'+\Delta P_{v_{n,j}})^2+(Q_{0_{n,j}}'+\Delta Q_{v_{n,j}})^2} \leq S_{ \mathrm{GFM}_{n,j}}.
\end{equation}
where $Q_{0_{n,j}}'$ relates to the voltage sag level and can be calculated using the formula $Q_{0_{n,j}}' = \frac {V_{t_{n,j}}}{V_{e_{n,j}}}Q_{0_{n,j}}$ since the reactive power output in steady states is express by $Q_{0_{n,j}} = V_{e_{n,j}}I_{r0_{n,j}}$. A worst-case scenario for ${V_{t_{n,j}}}$ is selected in the following case study for conservatism. Clearly, in the event of a severe fault, the active power output would decrease to zero and the GFM would solely generate reactive power to the grid to maintain the terminal voltage. Substituting \eqref{ReactivePowerInjectionGridCode} and \eqref{ReactivePowerTransient} into \eqref{ReactivePowerConstraints} while ensuring zero active power output leads to the worst-case scenario for the reactive power constraint: 
\begin{equation}
\label{ReactivePowerWorstCase}
           {V_{t_{n,j}}I_{r0_{n,j}}} -dV_{t_{n,j}}^2+dV_{e_{n,j}}V_{t_{n,j}} \leq S_{\mathrm{GFM}_{n,j}}.
\end{equation}
It should be noted that the derivation is only applicable when the value of $(Q_{0_{n,j}}'+\Delta Q_{v_{n,j}})$ is positive. This condition is met during a voltage sag or swell event. The worst-case transit voltage ${V_{t_{n,j}}}'$ is obtained by setting the derivative of the left hand side of \eqref{ReactivePowerWorstCase} with respect to ${V_{t_{n,j}}}$, to zero: 
\begin{equation}
\label{ReactivePowerWorstCase2}
           {V_{t_{n,j}}}' = \frac {I_{r0_{n,j}} + d{V_{e_{n,j}}}}{2d}.
\end{equation}
Therefore, the reactive power constraint can be rewritten by incorporating \eqref{ReactivePowerWorstCase2} into \eqref{ReactivePowerWorstCase}: 
\begin{equation}
\label{ReactivePowerWorstCase3}
           \frac{1}{4d} {I_{r0_{n,j}}^2} + \frac{V_{e_{n,j}}}{2} I_{r0_{n,j}} + \frac{d{V_{e_{n,j}}}^2}{4} \leq S_{\mathrm{GFM}_{n,j}}.
\end{equation}
Since $d$ is the droop gain controlling the reactive power injection during the fault event and $V_{e_{n,j}}$ represents the terminal voltage, expression \eqref{ReactivePowerWorstCase3} can be formulated as a SOC form and incorporated into the MISOCP-based SUC problem: 
\begin{equation}
\label{ReactivePowerWorstCaseSOC}
           \left( I_{r0_{n,j}}+dV_{e_{n,j}}\right) ^2 \leq 4dS_{\mathrm{GFM}_{n,j}},
\end{equation}
where the voltage $V_{e_{n,j}}$ is considered to be approximately equal to its nominal value.

Unlike conventional generators, inverter-based power electronic devices must comply with current limit constraints. The transient current limitations during the voltage sag events can be expressed as follows: 
\begin{equation}
\label{Currentlimit}
            \sqrt{\left(I_{a0_{n,j}}+\Delta I_{a_{n,j}}\right)^2+\left(I_{r0_{n,j}}+\Delta I_{r_{n,j}}\right)^2} \leq I_\mathrm{limit}.
\end{equation}
This constraint prevents IBR overcurrent and potential component damage. In a worst-case scenario where a severe voltage drop event is detected, the GFM will not provide any active power and all capacity will be used for reactive power injection, meaning no active current flows through the converter, i.e., $I_{a0_{n,j}}+\Delta I_{a_{n,j}}=0$. Combined with the droop control of reactive current injection in equation \eqref{ReactiveCurrentInjection}, the current limit constraint can be represented as follows: 
\begin{equation}
\label{Currentlimit2}
            I_{r0_{n,j}}+ {d(V_{e_{n,j}}-V_{t_{n,j}})} \leq I_\mathrm{limit}.
\end{equation}
Based on the constraints derived in \eqref{ReactivePowerWorstCaseSOC} and \eqref{Currentlimit2}, the reactive power output at steady-state, satisfying the relationship that $Q_{0_{n,j}} = V_{e_{n,j}}I_{r0_{n,j}}$, is therefore restricted.

\section{System Stability Constraints} \label{sec:3}
To guarantee the stable operation of large-scale power systems, extensive research has been conducted on system stability. Researchers have derived and implemented power system stability constraints, such as frequency stability \cite{9475967}, voltage stability \cite{9786660} and rotor angle stability \cite{9281042}, through optimal scheduling algorithms. Some of the security constraints are embedded in the software-defined grid scheduling model \cite{chu2023scheduling}. Besides, it has been widely investigated that integrating the GFM into the power system will improve the frequency stability and small-signal stability due to the inertia and grid strength provision. The proper amount of GFM in the system is therefore determined by GFM-related frequency stability constraints and small-signal stability constraints, which are presented in this section.

% However, the implementation of GFM introduces power and current headroom constraints as discussed in previous section, which may potentially lead to economic loss during operation. Therefore, it is desired to 

\subsection{System Transient Frequency Stability Constraints}
Based on the theory of the Centre-of-Inertia (CoI) model, frequency dynamics of a multi-machine power system can be expressed as a single swing equation \cite{7115982}:
\begin{equation}
    \label{sw1}
    2H\frac{\partial\Delta f(t)}{\partial t} = -D \Delta f(t) + \Delta R(t) -\Delta P_L,
\end{equation}
where $\Delta P_L$ represents the loss of generation at $t=0$ and can be regarded as a step disturbance. $D$ and $H$ denote the system damping and inertia. The Primary Frequency Response (PFR) from SGs $\Delta R(t)$ can be represented according to the following scheme \cite{6714513}, where $T_d$ is the delivery time of PFR:
\begin{equation}
\label{R}
\Delta R(t)=
     \begin{cases}
       \frac{R}{T_d}t &, \; 0\le t< T_d \\ 
       R &, \; T_d\le t
     \end{cases}.
\end{equation}
As aforementioned, wind turbines operating under grid-forming control mode possess frequency support functions including synthetic inertia provision. Consequently, the total available system inertia can be determined by summing the total inertia of conventional generators, denoted as $H_c$, and the GFM synthetic inertia provision from each GFM wind turbine among all the wind farms:
\begin{equation}
    \label{Htotal}
    H = H_c + {H}_{\mathrm{GFM}},
\end{equation}
where
\begin{equation}
    \label{H1}
    H_c =\frac{\sum_{g\in \mathcal{G}} H_g  P_g^\mathrm{max} N_g^\mathrm{up}}{f_0},
\end{equation}
with $H_g$ and $P_g^\mathrm{max}$ representing the inertia constant and installed capacity of the SG unit, respectively. $N_g^\mathrm{up}$ signifies the total number of online SGs, and $f_0$ denotes the nominal frequency. And the system GFM synthetic inertia contribution can be calculated as follows:
\begin{equation}
    \label{HGFM}
    {H}_{\mathrm{GFM}}= \sum_{j\in \mathcal{F}}\sum_{n\in \mathcal{N}} H_{\mathrm{GFM}_{n,j}}.
\end{equation}
% Wind turbines under grid-forming control mode have frequency support functions such as synthetic inertia provision. The total available system inertia can be obtained by adding the GFM synthetic inertia provision from each GFM wind turbine among all the wind farms:
% \begin{equation}
%     \label{Htotal}
%     H ={H}_{\mathrm{GFM}}= H_c+ \sum_{j\in \mathcal{F}}\sum_{n\in \mathcal{N}} H_{\mathrm{GFM}_{n,j}}.
% \end{equation}
% Incorporating the SI contribution from GFM into \eqref{sw1} yields:
% \begin{equation}
%     \label{sw2}
%     2\Big({H_c+ {H}_{\mathrm{GFM}}}\Big)\frac{\partial\Delta f(t)}{\partial t} = -D \Delta f(t) + \Delta R(t) -\Delta P_L.
% \end{equation}
Based on the frequency dynamics, the mathematical expressions for maximum instantaneous RoCoF can be derived according to the relationship $\Delta \dot f_\mathrm{max}\equiv\Delta \dot f|_{t=0^+}$: 
\begin{equation}
\label{rocof,fss}
    \Delta \dot f|_{t=0^+} = -\frac{\Delta P_L}{2\Big({H_c+ \sum_{j\in \mathcal{F}}\sum_{n\in \mathcal{N}} H_{\mathrm{GFM}_{n,j}}}\Big)}.
\end{equation}
The RoCoF limit constraint can be satisfied by adjusting the appropriate system inertia term $H$ with a predetermined magnitude of disturbance $\Delta P_L$. Moreover, the steady-state frequency deviation in premise of  $\Delta f_\mathrm{max}^\mathrm{ss}\equiv\Delta f|_{t=\infty}$ is obtained:
\begin{equation}
\label{nadir,fss}
    \Delta f|_{t=\infty} = \frac{R-\Delta P_L}{D}.
\end{equation}
From \eqref{nadir,fss}, it can be deduced that the PFR term $R$ plays a crucial role in maintaining steady state frequency deviation within acceptable bounds. Furthermore, the time-domain solution for frequency deviation can be derived by substituting \eqref{R} into \eqref{sw1} as follows:
\begin{equation}
\label{f(t)}
    \Delta f(t) = \left(\frac{\Delta P_L}{D}+\frac{2HR}{T_d D^2}\right)\left(e^{-\frac{D}{2H}t}-1\right) + \frac{R}{T_d D}t,
\end{equation}
valid $\forall t\in [0,t_n]$. In order to determine the time instance $t_n$ of frequency nadir, the solution can be obtained by setting the derivative of \eqref{f(t)} to zero, yielding:
\begin{equation}
\label{tn}
    \Delta \dot f(t_n)=0 \longmapsto t_n = \frac{2H}{D}\ln{\left(\frac{T_d D \Delta P_L}{2HR}+1\right)}.
\end{equation}
The frequency nadir $(\Delta f_\mathrm{max}\equiv\Delta f(t_n))$ can then be derived by combining \eqref{f(t)} with \eqref{tn} and yields the expression: 
\begin{equation}
\label{nadir}
    \Delta f(t_n) = \frac{2HR}{T_d D^2} \ln{\left(\frac{T_d D \Delta P_L}{2HR}+1\right)}-\frac{\Delta P_L}{D}.
\end{equation}
It is important to note that in order to ensure frequency stability, the frequency nadir must occur before $T_d$, i.e., $t_n\le T_d$. Utilizing the simplification algorithm proposed in \cite{8667397,9066910}, load damping is initially disregarded to eliminate exponential functions while preserving the constraint’s conservatism. Subsequently, system damping is approximately reformulated as a linear term. As a result, the frequency nadir constraint $\Delta f(t_n)\leq \Delta f_\mathrm{lim}$ can be expressed as follows:
\begin{equation}
\label{nadir_c}
    \Big(H_c+ \sum_{j\in \mathcal{F}}\sum_{n\in \mathcal{N}} H_{\mathrm{GFM}_{n,j}}\Big)R\ge \frac{\Delta P_L^2T_d}{4\Delta f_\mathrm{lim}}-\frac{\Delta P_L T_d }{4} D.
\end{equation}
Expressions \eqref{rocof,fss} and \eqref{nadir_c} demonstrate that maximum RoCoF and frequency nadir are dependent on the conventional inertia from SGs and synthetic inertia from GFMs, while the nadir constraint contains a highly nonlinear function. The frequency nadir constraint can be solved using either MISOCP \cite{9786660} or Mixed-Integer Linear Programming (MILP) based on the linearization approach proposed in \cite{9066910}, with the latter being chosen in this paper for its high accuracy and fast computation time. The potential for GFMs to provide synthetic inertia for frequency regulation can be observed in terms of economic saving, which is further optimized by the software-defined control.

\subsection{Formulation and Linearization of Small-signal Stability Constraints} 
In order to better demonstrate the constraints associated with system strength, it is necessary to address the formulation of the system susceptance matrix. A general power network can be considered as a set of wind farm nodes $(1\sim n)$, interior nodes $(n+1\sim n+m)$ and generator nodes $(n+m+1\sim n+m+k)$. The susceptance matrix of the network is defined by $\mathbf{B} \in \mathbb{R}^{(n+m+k)\times(n+m+k)}$, which can be expressed as: 
\begin{equation}
\label{B0+Bg}
            \mathbf{B}=\mathbf{B}^0+\mathbf{B}^g,
\end{equation}
where $\mathbf{B}^0$ is the matrix for transmission lines and the $\mathbf{B}^g$ denotes the additional increment for the reactance of SGs $X_g$, which is hereby defined in accordance with the state of generators $x_g$:
\begin{equation}
\label{B^g}
    B^g_{il}=
\begin{cases}
\frac{1}{X_{g}}x_g & \mbox{if $i=l \wedge \exists g \in \mathcal G$} ,\\
0  &\mbox{otherwise}.
\end{cases}
\end{equation}
$x_g$, $\forall g \in \mathcal G$ is the binary decision variable representing the on or off state for each generator. By applying the Kron reduction, only the dimension of wind farm nodes is expressed by the reduced susceptance matrix: 
\begin{equation}
\label{BReduced}
        \mathbf{B}_r=\mathbf{B}_1-\mathbf{B}_2\mathbf{B}_4^{-1}\mathbf{B}_3, 
\end{equation}
where
\begin{equation}
\label{B}
 \mathbf{B}=
\left[
\begin{matrix}
           \mathbf{B}_1\in \mathbb{R}^{n\times n} & \mathbf{B}_2\in \mathbb{R}^{n\times (m+k)} \\
           \mathbf{B}_3\in \mathbb{R}^{(m+k)\times n} &  \mathbf{B}_4\in \mathbb{R}^{(m+k)\times (m+k)}
           
  \end{matrix}   
  \right].
\end{equation}
Based on \cite{xin2023gridforming}, the generalized Short Circuit Ratio (gSCR) should always be maintained above the Critical gSCR (CgSCR) to ensure the small-signal stability of the system. Taking the GFM into account, the gSCR is defined by the minimum eigenvalue of the modified admittance matrix $Y_{eq}$: 
\begin{equation}
\label{gSCR0}
            gSCR = \min \lambda (Y_{eq}) \geq CgSCR,
\end{equation}
where 
\begin{equation}
\label{gSCR}
            Y_{eq} = \mathbf{S_B^{-1}}\left(\mathbf{B_r}+\mathbf{A} Y_{local}\right),
\end{equation}
with $\mathbf{S_B}= \mathbf{diag}\left(S_{B_1},...,S_{B_j} \right)$ being the diagonal matrix of the GFL capacity of each wind farm. The GFL capacity is unified by dividing a global base capacity, i.e., $S_{B_j}=\sum_{n\in \mathcal{N}}S_{GFL_{n,j}}/S_{global}$. $\mathbf{A} = \mathbf{diag}\left(\alpha_1,...,\alpha_j \right)$ is the diagonal matrix of the GFM penetration level for each wind farm. It should be noted that the penetration level of GFM in the $j$-th wind farm is defined by the ratio of the total capacity of grid-forming controlled wind turbines in the wind farm to its total rated capacity, as follows:
\begin{align}
\label{alpha}
            &\alpha_j = \frac{\sum_{n\in \mathcal{N}} S_{\mathrm{GFM}_{n,j}}}{S_{N_j}}, 
\end{align}
where ${S_{N_j}}$ denotes the rated capacity of the $j$-th wind farm. $Y_{local}$ is a parameter representing the per-unit susceptance between GFM and the external grid, which includes the GFM internal impedance\cite{xin2023gridforming}. It can be noticed that the software-defined decision variable $\alpha_j$ also appears in $\mathbf{S_B}$ determining the capacity of GFL since $S_{B_j}=S_{N_{j}}\left(1-\alpha_j\right)/S_{global}$. Therefore, the small-signal stability constraint \eqref{gSCR0} can be maintained by properly setting the operation status of SGs $x_g$, which is incorporated in the reduced susceptance matrix $\mathbf{B_r}$, and the GFM penetration level $\alpha_j \in [0,1]$ that determines both $\mathbf{S_B}$ and $\mathbf{A}$ matrices. 

However, the matrix constraint in \eqref{gSCR0} can not be incorporated into a MISOCP-based UC model directly since it contains decision-dependent eigenvalue operation that is difficult to deal with in an optimization problem in general. The data-driven approach proposed in \cite{9329077} is adopted for guaranteed accuracy, conservativeness and computational efficiency. Since \eqref{gSCR0} contains $\lvert \mathcal G \rvert + \lvert \mathcal F \rvert$ decision variable ($x_g$ and $\alpha_j$), the linearization of the SSC constraint can be formulated as follows:
\begin{equation}
\label{SSC_L}
             gSCR_{L} = \sum _{g\in \mathcal {G}} k_{g} x_g+ \sum _{j\in \mathcal {F}} k_{j} \alpha _j+ \sum _{m\in \mathcal {M}} k_{m} \eta _m \\  \geq CgSCR,
\end{equation}
where $k_{g}$, $k_{j}$ and $k_{m}$ are linear coefficients and $gSCR_{L}$ represents the linearized gSCR. In order to consider the original nonlinearity in \eqref{gSCR}, the term $k_{m} \eta _m$ is introduced to describe the interactions between every two units, which is defined as follows: 
\begin{equation}
\label{SSC_Units}
             \eta _{m} = {\begin{cases}x_{g_{1}}x_{g_{2}},\quad \text{if} \,\,g_{1},g_{2} \in \mathcal {G}\\ x_{g_{1}}\alpha _{g_{2}},\quad \text{if} \,\,g_{1}\in \mathcal {G},\,g_{2} \in \mathcal {F}\\ \alpha _{g_{1}}\alpha _{g_{2}},\quad \text{if} \,\,g_{1},g_{2} \in \mathcal {F} \end{cases}},\,\,\forall m \in \mathcal {M},
\end{equation}
where $ m\in \mathcal {M} =\lbrace g_1,\,g_2 \mid g_1,\,g_2\in \mathcal {G}\cup \mathcal{F} \rbrace$. It should be noted that including the 2nd-order terms results in an accurate linearization while the higher-order terms are neglected to balance accuracy and computational time. The linear coefficients $\mathcal {K} = \lbrace k_{g}, k_{j}, k_{m} \rbrace, \forall g,j,m$ are determined by solving the minimization problem: 
% \begin{subequations}
% \begin{align}
% \label{SSC_min}
%              &\min _{\mathcal {K}} \quad \sum _{\omega \in \Omega } \left(gSCR_{L}^{(\omega)} -gSCR^{(\omega)} \right)^2, \\ 
%              &\left. gSCR_{L}^{(\omega)} = gSCR_{L} \right|_{x_g^{(\omega)},\,\alpha _j^{(\omega)},\,\eta _m^{(\omega)}},    
% \end{align}
% \end{subequations}
\begin{subequations}
\label{K_OPT}
\begin{align}
\label{SSC_ctr2}
\min _{\mathcal {K}} &\quad \sum _{\omega \in \Omega _2} \left(gSCR^{(\omega)}- gSCR_{L}^{(\omega)} \right)^2, \\ 
&\quad \left. gSCR_{L}^{(\omega)} = gSCR_{L} \right|_{x_g^{(\omega)},\,\alpha _j^{(\omega)},\,\eta _m^{(\omega)}},\\
\text{s.t.} &\quad gSCR_{L}^{(\omega)}< CgSCR ,\,\,\forall \omega \in \Omega _1, \\ &\quad gSCR_{L}^{(\omega)}\geq CgSCR,\,\,\forall \omega \in \Omega _3,
\end{align}
\end{subequations}
where $\omega = \lbrace {x_g^{(\omega)},\,\alpha _j^{(\omega)},\,gSCR^{(\omega)}} \rbrace \in \Omega$ symbolizes the data set, which is generated based on the representative system conditions when evaluating the gSCR. The data set includes all the possible commitment combinations of SGs. However, for the continuous variable $\alpha_j \in [0,1]$, the finite data is obtained by evenly dividing the interval into $n_j$ regions. 
% Note that the coefficients used in \eqref{SSC_L} are sufficient to describe the correlation for all operating conditions with the size of the data set $\Omega$ being $2^{\lvert \mathcal G \rvert} \cdot n_j^{\lvert \mathcal F \rvert}$.
Moreover, $\Omega_1$, $\Omega_2$ and $\Omega_3$ are the subsets of $\Omega$, being defined as follows:
\begin{subequations}
\label{Omega123}
\begin{align}
 \Omega &= \Omega _1 \bigcup \Omega _2\bigcup \Omega _3,  \\ 
 \Omega _1  &= \left\lbrace \omega \in \Omega \mid gSCR^{(\omega)}< CgSCR \right\rbrace,  \\ 
 \Omega _2  &= \left\lbrace \omega \in \Omega \mid CgSCR \leq gSCR^{(\omega)}< CgSCR + \nu \right\rbrace,  \\ 
 \Omega _3  &= \left\lbrace \omega \in \Omega \mid CgSCR + \nu \leq gSCR^{(\omega)} \right\rbrace.  
\end{align}
\end{subequations}
The parameter $\nu \in \mathbb{R}$ is introduced to avoid infeasibility, and it should be chosen as small as possible for better performance. Remarkably, only the errors of data points in $\Omega_2$ are penalized in the objective function \eqref{SSC_ctr2}, since for the data points in $\Omega_1$ and $\Omega_3$, as long as they are classified on the correct side of the limits, the errors are of no concern. As a result, the proposed method, \eqref{K_OPT} and \eqref{Omega123}, ensures accurate approximation within a narrow region around the limits ($\Omega_2$) whereas the regression errors of the data points in the other two regions are insignificant.

% In order to guarantee the conservativeness that satisfies the relationship $gSCR_{L}^{(\omega)} \leq gSCR^{(\omega)}, \forall\omega$, the minimization problem can be modified as: 
% \begin{subequations}
% \begin{align}
% \label{SSC_ctr}
% \min _{\mathcal {K}} &\quad \sum _{\omega \in \Omega } \left( gSCR^{(\omega)}-gSCR_{L}^{(\omega)} \right), \\
% \text{s.t.}&\quad gSCR_{L}^{(\omega)} \leq gSCR^{(\omega)},\,\,\forall \omega \in \Omega. 
% \end{align}
% \end{subequations}

% Therefore, the linearized gSCR is always below the real one. Moreover, the modified classification problem is utilized to deal with the over-conservativeness \cite{9329077}: 

% where 

\section{Overall System Operation Considering GFM} \label{sec:4}
In power system optimal scheduling, the objective is to minimize the total cost while satisfying the relevant constraints. In this paper, a stochastic UC model previously developed in reference \cite{7115982} is employed. The uncertainty of wind power and generation outages are represented by constructing a suitable scenario tree, while the SUC problem minimizes the expected cost over all nodes in the given scenario tree:
\begin{equation}
    \label{eq:SUC}
    \min \sum _{s\in \mathcal {S}} \pi (s) \left(\sum _{g\in \mathcal {G}} C_g(s) + \Delta t(s) c^{sh} P^{sh}(s) \right).
\end{equation}
Note that $\pi (s)$ is the probability of scenario $s\in \mathcal {S}$ and $C_g(s)$ is the operation cost of the generation unit including start-up, marginal and no-load cost. The terms in load shedding cost $\Delta t(s) c^{sh} P^{sh}(s)$ denote the time step of scenario s, load shedding cost and the shed load, respectively. 

Several system operational constraints are taken into account, including power balance, thermal unit operation, GFM-related RoCoF \eqref{rocof,fss} and frequency nadir constraints \eqref{nadir_c}. For the novel GFM operational constraints, the maximum steady-state active power defined in equation \eqref{Pa2} and \eqref{ActivePowerConstraints2} is considered and the reactive power and current limit constraints expressed in \eqref{ReactivePowerWorstCaseSOC} and \eqref{Currentlimit2} are adopted. In addition, the linearized SSC derived in \eqref{SSC_L} is also included. More details regarding the model and SUC formulation can be found in \cite{7115982}. 

\section{Case Studies}\label{sec:5}
\begin{figure}[b]
        \vspace{-0.4cm}
    \centering
	\scalebox{0.2}{\includegraphics[]{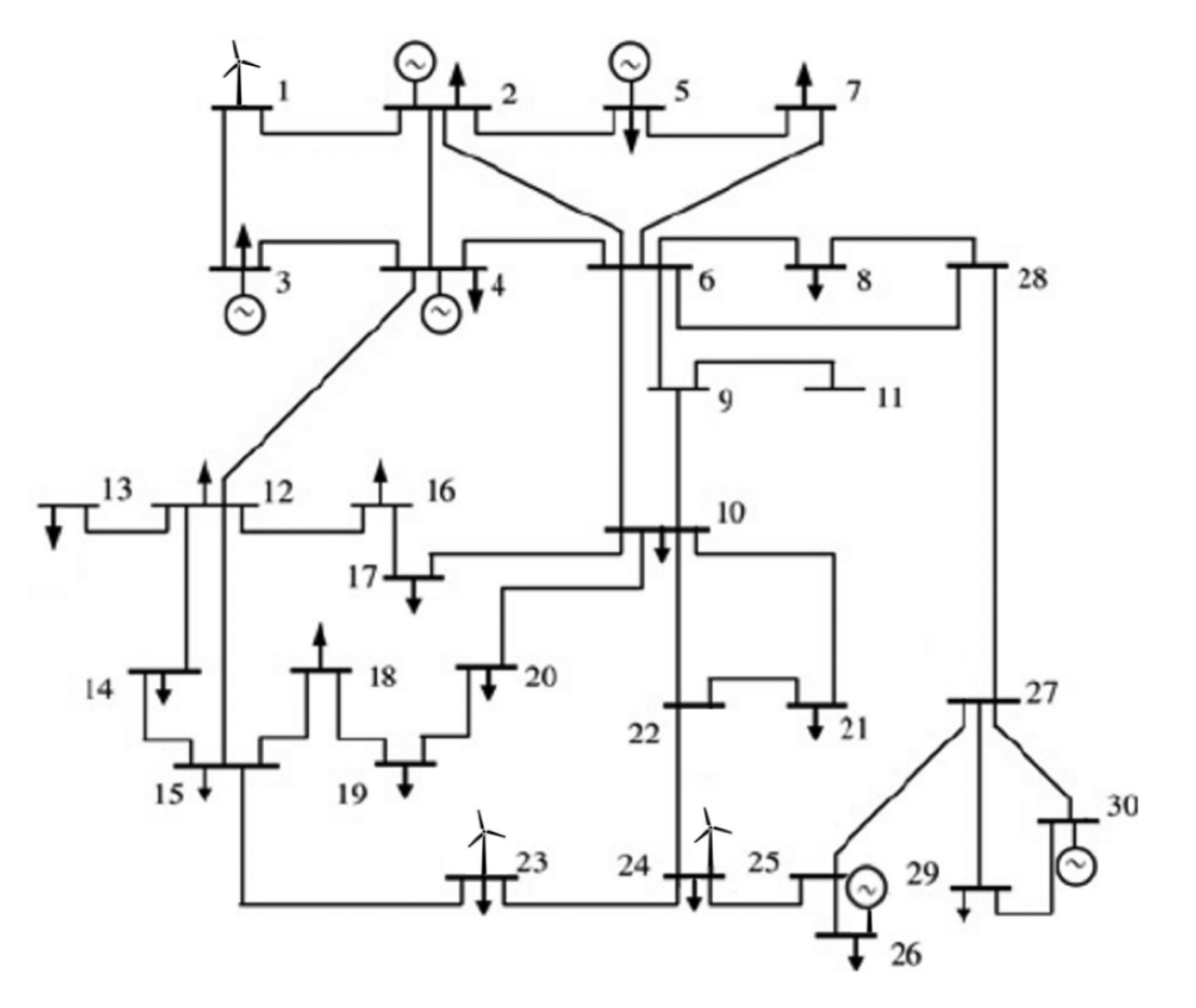}}
    \caption{\label{fig:IEEE 30BUS}Modified IEEE 30-bus power system.}

\end{figure}

Based on a modified IEEE 30-bus system, case studies have been implemented to illustrate the effectiveness of the proposed dynamic GFM penetration optimisation algorithm by solving the stability-constrained unit commitment problem while considering the reserve constraints. The topology of the modified system can be observed in Fig.~\ref{fig:IEEE 30BUS}, where SGs locate at buses 2, 3, 4, 5, 26, 30 and wind farms at buses 1, 23, 24. It is assumed that wind turbines in each wind farm are connected to the external grid via power electronic converters and the control methodology can switch between grid-following control and grid-forming control upon receiving commands from the system operator or the wind farm owner. The network data is referred from \cite{IEEE30BUS}, and the overall MISOCP-based unit commitment problem is solved by FICO Xpress on a laptop with an Intel® Core™ i7-11370H CPU @ 3.30GHz and 16 GB RAM.

The optimisation problem is solved in a horizon of 24 hours with the minimum time step being 1 hour. The system parameters are set as follows: base MVA $S_B=100$ MVA, load demand $P_D \in [230,620]$ MW, damping $D=0.5 \% P_D/1$ Hz, primary frequency response delivery time $T_d=10$ s and maximum power loss $\Delta P_L=50$ MW. The frequency limits of steady-state value, nadir and RoCoF set by National Grid are: $\Delta f_{\mathrm{lim}}^{\mathrm{ss}}=0.5$ Hz, $\Delta f_{\mathrm{lim}}=0.8$ Hz and $\Delta \dot{f}_{\mathrm{lim}}=0.5$ Hz/s. The GFM penetration level is regarded as a continuous number between 0 and 1 for each individual wind farm in the following case studies.

\subsection{GFL and GFM Switching Validation}
\begin{figure}[b]
    \vspace{-0.4cm}
    \centering

    \scalebox{1.19}{\includegraphics[]{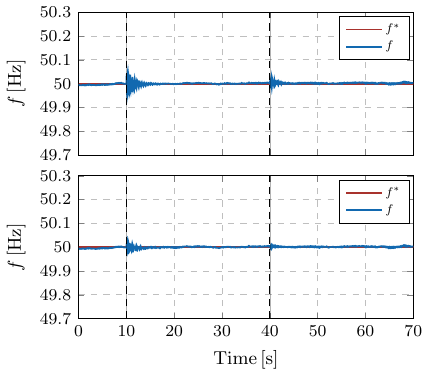}}    \caption{\label{fig:GFMGFL_Switching}Frequency transient performances with different GFM penetrations during the switching process between GFM and GFL.}

\end{figure}
Dynamic GFM allocation in the software-defined power grid needs to ensure smooth transient performance during the switching process between GFL and GFM. This subsection investigates such performance through a dynamic simulation of the transient process implemented in Matlab/Simulink. The GFL frequency is regulated by a PLL, with the converter functioning as a grid-feeding converter to inject active power into the system. In contrast, the GFM operates without PLL and provides ancillary service to the system. Fig.~\ref{fig:GFMGFL_Switching} depicts the transient frequency behaviour during the switching process. The figure below represents a switching process with a reduced switched GFM capacity of 50\% in the wind farm, in contrast to the figure shown above. It should be noted that a small, varying disturbance, intended to simulate real-world system operation conditions, has been incorporated into the simulation. During the initial $[0,10]\,\mathrm{s}$, the converter operates under grid-following control. At $t=10\,\mathrm{s}$, the GFL switches to GFM and then back to GFL at $t=40\,\mathrm{s}$. As indicated by the blue curves, the overall disturbance at switching points is relatively low, thereby demonstrating the feasibility of the proposed framework. Furthermore, a comparison of the two figures reveals that the system with reduced GFM penetration exhibits a smaller frequency deviation during the switching process.It should be noted that the impact of the transition can be further controlled by restricting the amount of simultaneous switching at one time.

\subsection{Small-Signal Stability Constraint Validation}

\begin{figure}[t]

    \centering
    \scalebox{1.12}{\includegraphics[]{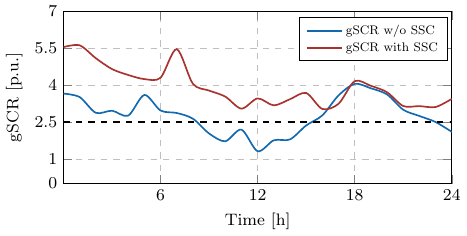}}
    \caption{\label{fig:SSC1}System strength with and without small-signal stability constraint.}
            \vspace{-0.4cm}
\end{figure}

% \begin{figure}[tb]
%     \centering
%     \scalebox{1.12}{\includegraphics[]{GFM_Figures/Figure_GFMUnequal/24H_NetDemand.pdf}}
%     \caption{\label{fig:SSC3}System net demand with and without small-signal stability constraint.}
% \end{figure}

This section presents the results validating the small-signal stability constraint. Fig.~\ref{fig:SSC1} illustrates the change in system strength with and without the SSC during a 24-hour simulation. The system strength is calculated using gSCR, which is dependent on the commitment status of SGs and the allocation of GFM and GFL at each hour. As shown in Fig.~\ref{fig:SSC1}, within the 24-hour time frame, the system strength is below the threshold during $[10,15]\,\mathrm{h}$ and the $24^{th}$ hour if no SSC is applied to the SUC problem (blue curve). However, with the inclusion of the SSC (red curve), it can be observed that SSC can successfully improve system gSCR and ensure a small-signal stable system by optimizing the SG status and the GFM penetration level to meet the grid strength requirements.

The variations of GFM penetration level for each individual wind farm (Fig.~\ref{fig:SSC2} above) and the system net demand (Fig.~\ref{fig:SSC2} below) with and without the SSC are presented, demonstrating how the SSC improves the overall grid strength. The system net demand represents the active power output of SGs, which means the wind shed power, GFM reserved power and active wind power output are excluded. It can be viewed during the first 10 hours, when the system's available wind power is sufficient, which corresponds to the system net demand being extremely low, the GFM penetration level is relatively high. In this scenario, the system strength falls within the acceptable secure region due to the substantial contributions from a significant number of GFMs. Moreover, there is no discernible difference in $\alpha_j$ across various wind farms, regardless of the consideration of SSC. This is attributed to the consistency in inertia provision for wind farms situated in diverse locations. 

Additionally, when the available wind power decreases from a sufficient level to an intermediate level between 10 and 16 $\mathrm{h}$, the system strength may violate the SSC if not being confined and can be improved from two aspects. Firstly, the penetration level of GFM is increased, from the dashed curves to the solid curves, to directly enhance the system gSCR. It can be deduced that the GFMs in Wind Farm 1 are more effective in improving the system strength, as a greater number of wind turbines in this farm operate under grid-forming control compared to the other two wind farms, following the implementation of SSC. However, in this scenario where wind power is insufficient, an increase in GFM penetration results in more power being generated by SGs, meeting grid strength requirements at a non-negligible cost increase. Secondly, the slight increase in the system’s net demand also indicates that the number of online generators is increased to support the grid strength under these circumstances.
\begin{figure}[t]

    \centering
    \scalebox{1.12}{\includegraphics[]{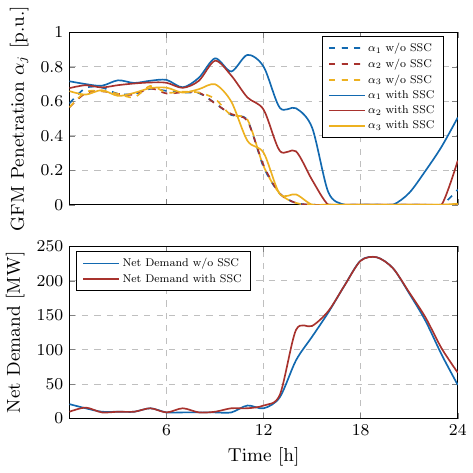}}
    \caption{\label{fig:SSC2}GFM penetration level $\alpha_j$ for the $j$-th wind farm and system net demand $P_N$ with and without small-signal stability constraint.}
            \vspace{-0.4cm}
\end{figure}
Furthermore, in situations where wind power is relatively low (during $[16,22]\,\mathrm{h}$), the additional power reserve requirements when operating under the GFM model cannot be satisfied, resulting in the optimal GFM penetration level to be a very low value. During these hours, the majority of active power is supplied by the SGs rather than wind farms, leading to sufficient system strength even without the inclusion of the SSC. As a result, from the perspective of the system operator, few adjustments to the GFM penetration level or SG commitment status are needed in this scenario.
\begin{figure}[b]
            \vspace{-0.4cm}
    \centering
    \scalebox{1.12}{\includegraphics[]{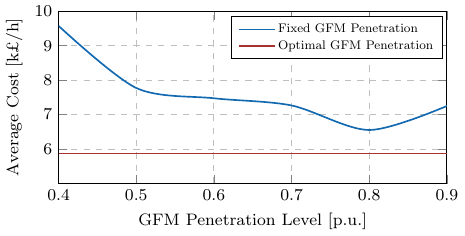}}
    \caption{\label{fig:GFM_COST}System operation cost with fixed or optimal GFM penetration control method.}

\end{figure}

\subsection{Value of Dynamic GFM Penetration Optimisation}

\begin{figure}[t]

    \centering
    \scalebox{1.20}{\includegraphics[]{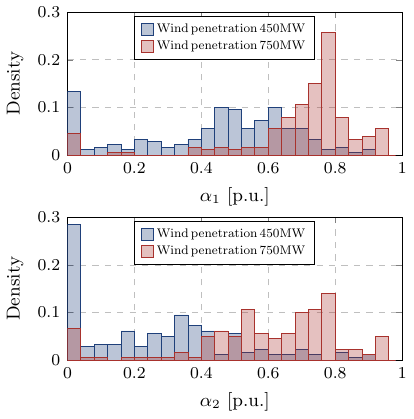}}\caption{\label{fig:GFM_Wind_Histogram} Optimal GFM penetration level histograms for Wind Farm 1 (above) and Wind Farm 2 (below) with different wind penetrations.}
        \vspace{-0.5cm}

\end{figure}
Based on the theoretical analysis in previous sections, GFMs can provide ancillary services and improve grid strength, but at the cost of deloaded operation for GFMs in steady states due to power reserve constraints. Despite the availability of day-ahead generation and demand forecasting from the system operator, most literature does not account for the dynamic modification of GFM penetration levels $\alpha_j$, leading to a loss in profit. In this section, we investigate and demonstrate the benefits of dynamic GFM penetration optimization. Fig.~\ref{fig:GFM_COST} displays the difference in average cost between fixed and optimal GFM penetration. The average cost is determined based on hourly operational costs.

With the fixed GFM penetration control (blue curve), the percentage of GFM in a wind farm remains constant over time, whilst the optimal GFM penetration (red curve) is generated based on the solution to the GFM-related power reserve and system stability constrained SUC problem. As shown in Fig.~\ref{fig:GFM_COST}, it is obvious that the average cost for optimal GFM penetration control ($\mathrm{5.87\, k\pounds/h}$) is consistently lower than that of the fixed GFM method (with the lowest average cost being $\mathrm{6.55\, k\pounds/h}$), demonstrating the superiority of the proposed approach. Moreover, for the fixed GFM penetration case (blue curve), it can be observed that the average cost is high when the GFM penetration level is set lower (approximately $\mathrm{9.57 \,k\pounds/h}$ when $\alpha_j = 0.4$). Lower GFM penetration results in an insufficient provision of synthetic inertia by the wind farm, as the GFMs are expected to provide more inertial service for the system. This lack of inertia conflicts with frequency nadir and RoCoF constraints, necessitating more SGs to be online to provide reserve or inertia to sustain frequency in the event of an unintentional occurrence, leading to higher average system operational costs. In this scenario, the cost of additional inertia provision by SGs outweighs the power reserve cost of GFMs. However, the cost begins to rise from $\alpha_j = 0.8$, which can be attributed to the GFM power reserve constraint. In conclusion, when there is sufficient system inertia and grid strength, increasing the number of GFMs results in reduced wind power generation, leading to higher operation costs.

\subsection{Impact of Wind Capacity on Optimal GFM Penetration and System Operation Cost}

\begin{figure}[b]
    \vspace{-0.5cm}
    \centering
    \scalebox{1.12}{\includegraphics[]{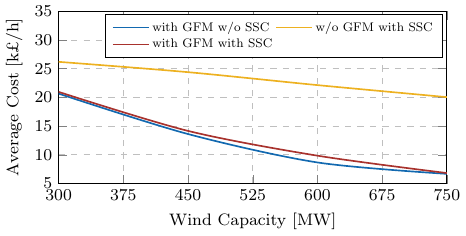}}\caption{\label{fig:Wind_cost}System operation cost for varying wind capacities under different constraints.}

\end{figure}

Given the significant impact of available wind power on system security and optimal scheduling, it is valuable to investigate the effect of wind capacity on optimal GFM allocation and system operational cost. Fig.~\ref{fig:GFM_Wind_Histogram} shows the density distributions of optimal GFM penetration levels for different wind penetrations in two wind farms. Focusing on the histogram of the first wind farm, as indicated by the blue bars, for a system wind penetration of $450\, \mathrm{MW}$, the optimal $\alpha_1$ is more concentrated between 0.4 to 0.6. In contrast, as represented by the red bars, when the wind capacity is $750\, \mathrm{MW}$, the majority of optimal GFM penetration falls within 0.6 and 0.9. It can be concluded that the optimal GFM rises with the increment in system wind penetration. With unchanged load demand, higher available wind power is more likely to result in a larger amount of wind power being curtailed for generation-demand balance in some time steps. However, when accounting for GFM-related power reserve constraints, wind shed power intended for curtailment can be considered as GFM power reserve, enabling GFMs to optimize their power output with zero generation loss due to headroom requirements. Thus the GFM operation constraints will not limit the upper bound of $\alpha_j$ and the average GFM penetration level will rise. Additionally, this fact is corroborated by cases where the optimal solution involves no GFMs in the wind farm (the bars furthest to the left, where $\alpha_j = 0$), and the density is lower for systems with high wind penetration. Furthermore, a comparison of the histograms for the two wind farms reveals that $\alpha_1$ generally distributes at a higher level than $\alpha_2$. This corroborates the conclusion presented in Part B that Wind Farm 1 is more effective in improving system strength. It also further proves that the location of a wind farm impacts the optimal allocation of GFM.

It is widely recognized that as wind capacity increases, system operational cost decreases due to the increased power provided by wind turbines. Hence, it is also valuable to compare system costs under different operational conditions for varying wind capacities. As illustrated by the curves in Fig.~\ref{fig:Wind_cost}, to maintain SSC, system operational cost is significantly high when all wind farms operate under GFL control (yellow curve). This suggests that all required inertia for frequency security and contributions to improving grid strength must be met by committing SGs, resulting in a substantial generation cost. With the implementation of GFM (red curve), the average operational cost reduces dramatically, demonstrating the effectiveness of the GFM on system frequency and strength support. On the other hand, this cost is higher compared to the case when there is no grid strength requirement (blue curve). This is because meeting the small-signal constraint involves either increasing the GFM penetration level or the online SG capacity, both of which result in an increase in average cost, as discussed in previous sections. 

Furthermore, it can be observed that the largest difference in average cost between cases with and without SSC occurs when wind capacity is approximately $600\, \mathrm{MW}$, and the cost difference is relatively small for systems with very large or small wind penetration power. This further supports the conclusion from Section V Part B that small-signal instability is more likely to occur when wind power is at an intermediate level, as in other circumstances the grid strength is either supported by SGs (in scenarios with insufficient wind) or by GFMs (in scenarios with sufficient wind).

    \vspace{-0.15cm}
\section{Conclusions} \label{sec:6}

This paper investigates a new research problem concerning the optimal allocation of GFM and GFL to balance the ability of GFMs to enhance grid strength against the potential economic loss resulting from reserved headroom in software-defined power grids. An optimization framework is proposed for the first time, to dynamically determine the optimal allocation of GFMs and GFLs in power systems at each time step according to system conditions, which ensures GFM-related operational constraints, system stability and minimum operational cost. Moreover, the optimal scheduling problem is solved and case studies performed on the modified IEEE 30-bus system reveal substantial economic advantages, as the optimal allocation of GFMs in the power system can be dynamically determined while ensuring power reserve and system stability constraints. It is also demonstrated that the optimal GFM allocation is influenced by both the penetration of wind power generation and the location of wind farms within the power system.

% References section
\bibliographystyle{IEEEtran}
    \vspace{-0.15cm}
\bibliography{bibliography}
\end{document}